%% file: struct.tex
\documentclass{article}
\usepackage{graphicx}
\usepackage[below]{placeins}
\usepackage{fancyvrb}
\usepackage{psfig}
\usepackage{epsf}

\newcommand{\beq}{\begin{equation}}
\newcommand{\eeq}{\end{equation}}
\newcommand{\beqa}{\begin{eqnarray}}
\newcommand{\eeqa}{\end{eqnarray}}

\renewcommand{\(}{\left(}

\newcommand{\Echain}{E_{\mbox{\scriptsize chain}}}
\newcommand{\lsheet}{\lambda_{\mbox{\scriptsize sheet}}}
\newcommand{\lhelix}{\lambda_{\mbox{\scriptsize helix}}}

\begin{document}

\begin{titlepage}

\begin{flushright}
LU TP 00-07\\
SLAC-PUB-8429\\
April 13, 2000
\end{flushright}

\vspace{.25in}

\LARGE

\begin{center}
{\bf A Novel Approach to Structure Alignment}\\
\vspace{.3in}
\large

Mattias Ohlsson,  
Carsten Peterson,
Markus Ringn\'er\footnote{\{mattias,carsten,markus\}@thep.lu.se}\\ 
\vspace{0.10in}
Complex Systems Division, Department of Theoretical Physics\\ 
University of Lund,  S\"{o}lvegatan 14A,  S-223 62 Lund, Sweden \\
{\tt http://www.thep.lu.se/complex/}

\vspace{0.15in}  
Richard Blankenbecler\footnote{rzbth@slac.stanford.edu}\\ 
\vspace{0.10in}
Stanford Linear Accelerator Center \\ 
P.O. Box 4349, Stanford, CA 94309, USA

\vspace{0.25in}

Submitted to {\it Structure with Folding \& Design}

\end{center}

\large

\vspace{0.8in}

{\it Running Head}: Structure Alignment of Proteins


\newpage

\begin{center}
{\bf Abstract}
\end{center}

{\bf Background:} Aligning protein structures is a highly relevant task.  It
enables the study of functional and ancestry relationships between
proteins and is very important for homology and threading methods in
structure prediction. Existing methods typically only partially
explore the space of possible alignments and being able to efficiently
handle permutations efficiently is rare.

{\bf Results:} A novel approach for structure alignment is presented, where
the key ingredients are: (1) An error function formulation of the
problem simultaneously in terms of binary (Potts) assignment variables
and real-valued atomic coordinates. (2) Minimization of the error
function by an iterative method, where in each iteration a mean field
method is employed for the assignment variables and exact
rotation/translation of atomic coordinates is performed, weighted with
the corresponding assignment variables. The approach allows for
extensive search of all possible alignments, including those involving
arbitrary permutations. The algorithm is implemented using a
$C_{\alpha}$-representation of the backbone and explored on different
protein structure categories using the Protein Data Bank ({\sc Pdb})
and is successfully compared with other algorithms.

{\bf Conclusions:} The approach performs very well with modest CPU
consumption and is robust with respect to choice of parameters.  It is
extremely generic and flexible and can handle additional
user-prescribed constraints easily. Furthermore, it allows for a
probabilistic interpretation of the results.

\end{titlepage}

\large

\newpage

\section*{Introduction}
\input{introduction}


\section*{Methods}
\input{methods}


\section*{Results}
\label{s:results}
\input{results}


\section*{Discussion}
\input{discussion}

\bibliographystyle{unsrt}
\bibliography{struct}

\end{document}

%% file: introduction.tex
Aligning protein structures is a subject of utmost relevance. 
It enables the study of functional relationship between proteins 
and is very important for homology and threading methods in structure 
prediction. Furthermore, by grouping protein structures into fold 
families and subsequent tree reconstruction, ancestry and evolutionary issues may get 
unraveled. 

Structure alignment amounts to matching two 3D structures such that 
potential common substructures, e.g. $\alpha$-helices, have priority. 
The latter is accomplished by allowing for gaps in either of the chains. 
Also, the possibility of permuting sites within a chain may be beneficial. 
At first sight, the problem may appear very similar to sequence 
alignment, as manifested in some of the vocabulary (gap costs etc.). 
However, from an algorithmic standpoint there is a major difference. 
Whereas sequence alignment can be solved within polynomial time using 
dynamical programming methods \cite{needle}, this is not the case 
for structure alignment since rigid bodies are to be matched. Hence, 
for all structure alignment algorithms the scope is limited to high 
quality approximate solutions. 

Existing methods for structure alignment fall into two broad classes, 
depending upon whether one (1) directly minimizes the {\it inter}-atomic 
distances between two structures or (2) minimizes the distance between 
substructures that are either preselected or supplied by an algorithm 
involving {\it intra}-atomic distances.

One approach within the first category is the iterative dynamical 
programming method \cite{laur,levitt}, where one first computes a 
distance matrix between all pairs of atoms (e.g. $C_{\alpha}$) forming 
a similarity matrix, which by dynamical programming methods gives rise to  
an assignment matrix mimicking the sequence alignment procedure. One of 
the chains is then moved towards the other by minimizing the distance 
between assigned pairs. This method does not allow for permutations.
Another inter-atomic approach is pursued in~\cite{cohen}, where the 
area rather than distances between two structures is minimized.

In~\cite{holm} the approach is different. Here one compares
distance matrices within each of the two structures to be aligned,
which provide information about similar substructures. The latter are
subsequently matched. A similar framework is used in~\cite{vast}
and also in~\cite{lu}. Not surprisingly,
in~\cite{holm,vast} and \cite{lu} permutations can in
principle be dealt with.

There are implementation issues shared by both methodologies above. 
One is structure encoding ($C_{\alpha}$ and/or $C_{\beta}$ of the 
chains). For many comparisons $C_{\alpha}$ appears to be sufficient, 
whereas in some cases $C_{\beta}$ is needed. Also, the choice of distance 
metric is a subject of concern in order to avoid the influence of 
outliers.

The iterative dynamical programming method \cite{levitt} has been
extensively assessed for backbone structures \cite{levitt3} from
the {\sc Scop}~\cite{scop} database, in which protein structures
have been classified by visual inspection. Some comparisons with {\sc
Scop} have also been performed \cite{matsuo} using the method
in~\cite{vast}.

Here we present a novel approach, which shares some of its philosophy 
from the iterative dynamical programming method \cite{levitt}. 
Its key ingredients are: (A) An error function formulation of the 
problem simultaneously in terms of binary (Potts) assignment variables 
and real-valued atomic coordinates and (B) minimization of the error 
function by an iterative method, where each iteration contains two steps:
\begin{enumerate}
\item A mean field procedure for minimizing with respect to  
the assignment variables.
\item Exact rotation and translation of atomic coordinates weighted with 
the corresponding assignment variables.
\end{enumerate}
The approach, which is very general, has some very appealing 
properties:
\begin{itemize}
\item Implicit complete exploration of the entire space of alignments, 
which allows for arbitrary permutations. To our knowledge, no other 
approach has this feature.
\item Probabilistic interpretation of the results. This feature is 
present without tedious Monte Carlo estimates since  
the algorithm is deterministic. Among other things, this implies 
that the approach is less sensitive to the choice of distance metric, 
since the distances are weighted with fuzzy numbers.
\item With its generality, almost arbitrary additional constraints are 
easily incorporated into the formalism including different functional 
forms of gap penalties.

\end{itemize}

The approach is tested using $C_{\alpha}$-representation of backbones, 
by comparing the results with the approaches of \cite{levitt} and 
\cite{holm} as implemented in the {\sc Yale Alignment Server}  
and {\sc Dali} respectively and in one instance also with~\cite{vast}
({\sc Entrez}). In choosing protein pairs to align we
followed~\cite{levitt3} to a large extent. In~\cite{levitt3} pairs
with marginal sequence overlap but where each protein in a pair
belongs to the same {\sc Scop} superfamily and therefore have a
similar structure were picked for assessment. We selected pairs from a
varied selection of the families used in~\cite{levitt3} to test our
algorithm:

\mbox{} {$\bullet$} {\it Dihydrofolate Reductases} ($\alpha/\beta$) \\ 
\mbox{} {$\bullet$} {\it Globins} (all-$\alpha$) \\
\mbox{} {$\bullet$} {\it Plastocyanin/azurin} (all-$\beta)$  \\
\mbox{} {$\bullet$} {\it Immunoglobulins} (all-$\beta)$ 

In addition, we test the permutation capacity  
of our approach, by aligning:

{$\bullet$} {\it Permuted proteins} (winged helix fold)

When assessing the algorithm, we limit ourselves to a core version,
where $C_{\beta}$ degrees of freedom are not included. Also, no
post-processing of the results is done. We defer such elaborations and
others to forthcoming publication. Nevertheless, the core version of
our approach is already very competitive even for chains, where
permutations are not needed. For the latter case, the other algorithms
could not be tested using the corresponding WWW-servers. In the
instances, where we have tested it for this kind of problems, it also
performs well.

The algorithm is implemented in $C$++. Given its generality and power, 
the CPU demand  is quite modest -- it scales like the chain lengths 
squared and on the average requires a few seconds on a Pentium 400MHz PC.

%% file: methods.tex
\subsection*{The Algorithm}
In what follows we have two proteins with $N_1$ and $N_2$ atoms 
to be structurally aligned. This is accomplished by a series of 
weighted rigid body transformations of the first chain, keeping 
the second chain fixed. We denote by $\mathbf{x}_i$ ($i=1,...,N_1$) 
and $\mathbf{y}_j$ ($j=1,...,N_2$) the atom coordinates of the first 
and second chain, respectively. The phrase "atom" will be used 
throughout this paper in a generic sense -- it could represent 
individual atoms but also groups of atoms. In our applications it  
will mean $C_{\alpha}$-atoms along the backbone.
A square distance metric between the chain atoms is used, 
\beq
\label{d2}
d_{ij}^2 = |\mathbf{x}_i - \mathbf{y}_j|^2 
\eeq
but the formalism is not confined to this choice.

We start by discussing the encodings and error function and then we
present a method for minimizing the latter.

{\bf The Gapless Case.} For pedagogical reasons, we start off with the
gapless case with $N_1 = N_2$. We define binary assignment
variables $s_{ij}$ such that $s_{ij} = 1$ if atom $i$ in one chain
matches $j$ in the other and $s_{ij} = 0$ otherwise. Since every atom
in one chain must match one atom in the other, the following
conditions must be fulfilled:

\beq
\label{cond1}
\sum_{i=1}^{N_1} s_{ij} = 1 \quad j=1,\ldots,N_2
\eeq
\beq 
\label{cond2}
\sum_{j=1}^{N_2} s_{ij} = 1 \quad i=1,\ldots,N_1
\eeq
A suitable error function to minimize subject to the above
constraints (Eqs.~(\ref{cond1},\ref{cond2})) is
\beq
\label{Echain}
  \Echain = \sum_{i=1}^{N_1}\sum_{j=1}^{N_2} s_{ij}d_{ij}^2 
\eeq
where the spatial degrees of freedom, $\mathbf{x}_i$, are contained in
the distance matrix $d_{ij}^2$. Thus whenever $s_{ij}$=$1$ one adds 
a penalty $d_{ij}^2$ to $\Echain$. Note that Eq. (\ref{Echain}) is to 
be minimized both with respect to the binary variables $s_{ij}$ and the
real-valued coordinates $\mathbf{x}_i$.

{\bf The Gapped Case.} Allowing for gaps in either of the chains is
implemented by extending $s_{ij}$ to include $0$-components in a
compact way; $s_{i0}=1$ and $s_{0j}=1$ if an atom ($i$~or~$j$) in one
chain is matched with a gap in the other and vice versa. Hence, gap
positions are not represented by individual elements in $s_{ij}$;
rather the gap-elements correspond to common sinks. The matrix
$\mathcal{S}$, with elements $s_{ij}$, containing gap-elements is
shown in Eq.~(\ref{S}).

\beq \quad \mathcal{S} = \left(
\begin{array}{l|llll}
           & s_{01}   & s_{02}   & ... &   s_{0N_2} \\
  \hline  
  s_{10}   & s_{11}   & s_{12}   & ... &   s_{1N_2} \\ 
  s_{20}   & s_{21}   & s_{22}   & ... &   s_{2N_2} \\ 
  .&&&& \\
  .&&&& \\
  .&&&& \\
  s_{N_10} & s_{N_11} & s_{N_12} & .... &  s_{N_1N_2} \\ 
\end{array}
\right)
\label{S}
\eeq

\begin{figure}[tbh]
  \parbox[t]{8cm}{
    \begin{center}
      \vspace{0.5cm}
      \includegraphics[angle=0,width=5cm]{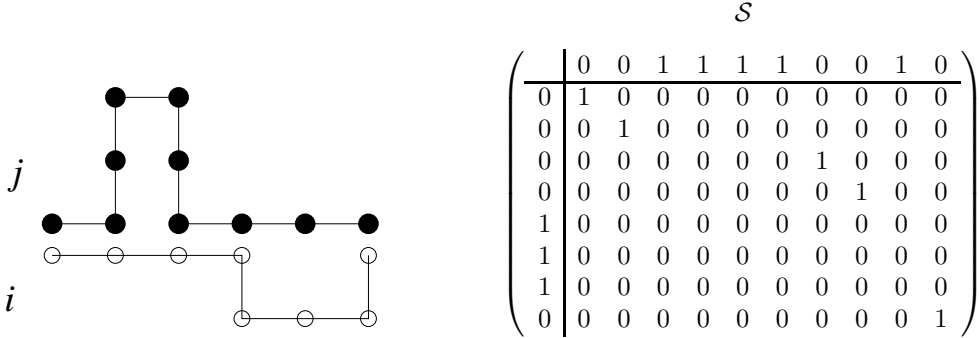}
    \end{center}
    }
  \parbox[t]{5cm}{
    \hspace{3.1cm}$\mathcal{S}$
    \[\left(
    \begin{array}{c|cccccccccc}
        & 0 & 0 & 1 & 1 & 1 & 1 & 0 & 0 & 1 & 0  \\
      \hline  
      0 & 1 & 0 & 0 & 0 & 0 & 0 & 0 & 0 & 0 & 0  \\
      0 & 0 & 1 & 0 & 0 & 0 & 0 & 0 & 0 & 0 & 0  \\
      0 & 0 & 0 & 0 & 0 & 0 & 0 & 1 & 0 & 0 & 0  \\
      0 & 0 & 0 & 0 & 0 & 0 & 0 & 0 & 1 & 0 & 0  \\
      1 & 0 & 0 & 0 & 0 & 0 & 0 & 0 & 0 & 0 & 0  \\
      1 & 0 & 0 & 0 & 0 & 0 & 0 & 0 & 0 & 0 & 0  \\
      1 & 0 & 0 & 0 & 0 & 0 & 0 & 0 & 0 & 0 & 0  \\
      0 & 0 & 0 & 0 & 0 & 0 & 0 & 0 & 0 & 0 & 1  \\
    \end{array}
    \right)\]
    }
  \caption{A simple example of the assignment matrix $\mathcal{S}$ (right) corresponding
    to the matching of the two toy chains (left).}
  \label{fig_gap}
\end{figure}

Some caution is needed when generalizing
Eqs.~(\ref{cond1},\ref{cond2}) to host gaps, since the 
elements of the first row
and column (gap-mappings containing the index $0$) in Eq. (\ref{S})
differ from the others in that they need not sum up to $1$. Hence
Eqs. (\ref{cond1},\ref{cond2}) becomes
\beqa
\label{cond1a}
\sum_{i=0}^{N_1} s_{ij} = 1; \quad j=1,\ldots,N_2 \nonumber \\
\label{cond2a}
\sum_{j=0}^{N_2} s_{ij} = 1; \quad i=1,\ldots,N_1 
\eeqa
where the first condition can be rewritten as
\beq
\label{cond1b}
\sum_{i=1}^{N_1} s_{ij} = 1 \quad \mbox{or} \quad
\sum_{i=1}^{N_1} s_{ij} = 0; \quad j=1,\ldots,N_2
\eeq
The encoding ($s_{ij}$) of matches and gaps is illustrated in 
Fig. \ref{fig_gap} with a simple example.

Assuming a constant penalty per inserted gap one has the error function 
\beq
\label{E2}
E = \Echain 
   + \sum_{i=1}^{N_1}\lambda^{(1)}_i s_{i0}
   + \sum_{j=1}^{N_2}\lambda^{(2)}_j s_{0j}
\eeq
where $\lambda^{(1)}_i$is the cost for matching atom $i$ in the first
chain with a gap in the second chain, and similarly for
$\lambda^{(2)}_j$. The position dependence of the gap costs,
$\lambda^{(1)}_i$ and $\lambda^{(2)}_j$, originates from the fact that
it is desirable not to break $\alpha$-helix and $\beta$-strand
structures.

In Eq. (\ref{E2}) the gap penalties are proportional to gap lengths.
In sequence alignment it is conjectured that gap penalties consist of
two parts; a penalty for opening a gap and then a penalty proportional
to the gap length.  As in \cite{levitt}, we will for structure
alignment here adopt the same gap cost philosophy,
i.e. $\lambda^{(1)}_i$ and $\lambda^{(2)}_j$ for opening a gap and a
position-independent $\delta$ per consecutive gap.
Hence, Eq. (\ref{E2}) generalizes to 
\beqa
\label{E3}
E &=&  \Echain + 
       \sum_{i=1}^{N_1}\lambda^{(1)}_i s_{i0} + 
       \sum_{j=1}^{N_2}\lambda^{(2)}_j s_{0j} \nonumber \\
  &+&  \sum_{i=2}^{N_1}\(\delta - \lambda^{(1)}_i\)s_{i-1,0}s_{i0} + 
       \sum_{j=2}^{N_2}\(\delta - \lambda^{(2)}_j\)s_{0,j-1}s_{0j}
\eeqa
where products like $s_{i-1,0}s_{i0}$ are 1 if two adjacent atoms 
are matched to gaps.

{\bf Minimization.} Next we need an efficient procedure for minimizing
$E$ with respect to both $s_{ij}$ and $\mathbf{x}_i$ subject to the constraints
in Eqs. (\ref{cond2a},\ref{cond1b}).  As mentioned above, this
minimization problem is non-trivial due to the rigid body
constraint. A similar problem in terms of fitting structures with 
relevance factors was probed in \cite{ohl} for track finding problems 
with a template approach using the mean field approximation. Here 
we will adopt a similar approach.

In our formulation, the inherent optimization difficulty resides in
the binary part ($s_{ij}$) of the problem. Hence, minimizing
Eq. (\ref{E3}) using a simple updating rule for $s_{ij}$ will very
likely yield poor solutions due to local minima. Well known stochastic
procedures such as simulated annealing ({\bf SA}) \cite{kirk} for
avoiding this are too costly from a computational standpoint. In the
{\it mean field} ({\bf MF}) approach \cite{klass}, the philosophy
behind SA is retained, but the tedious simulations are replaced by an
efficient deterministic process. The binary variables $s_{ij}$ are
then replaced by continuous \textit{mean field} variables $v_{ij} \in
[0,1]$, with a dynamics given by iteratively solving the MF equations
for a decreasing set of temperatures $T$ down to $T_0$, where most of
the $v_{ij}$ approach either $1$ or $0$. These continuous MF
variables can evolve in a space not accessible to the original
intermediate variables. The intermediate configurations at non-zero
$T$ have a natural probabilistic interpretation.

For $s_{ij}$ satisfying Eq. (\ref{cond2a}), the MF equations for the
corresponding $v_{ij}$ read
\beq
v_{ij} = \frac{e^{u_{ij}/T}}{\displaystyle \sum_{k=0}^{N_2} e^{u_{ik}/T}};
        \quad i=1,...,N_1
\label{MF}
\eeq
where the force $u_{ij}$ is given by
\beq
u_{ij} = -\frac{\partial E}{\partial v_{ij}}
\label{u}
\eeq
and is computed by substituting $s_{ij}$ with $v_{ij}$ in $E$ (Eq. (\ref{E3})). 
Note that the desired normalization condition, Eq. (\ref{cond2a}),
\beq
\sum_{j=0}^{N_2} v_{ij}=1; \quad i=1,...,N_1
\eeq
is fulfilled automatically in Eq. (\ref{MF}). The other condition
(Eq.(\ref{cond1b})) is enforced by adding a penalty term
\beqa
\label{EG1}
E_{\gamma}& = &
\gamma\sum_{j=1}^{N_2}[(\sum_{i=1}^{N_1}v_{ij})
(\sum_{k=1}^{N_1}v_{kj}-1)] \nonumber \\
 & = & \gamma\sum_{i=1}^{N_1}\sum_{k\neq i}^{N_1}\sum_{j=1}^{N_2} v_{ij}v_{kj}
\eeqa
where $\gamma$ is a parameter and the last equality follows from the 
fact that $v_{ij}^2$ = $v_{ij}$ for $T$=0. 

So far we have only looked at the assignment part when minimizing the
error function. When updating the mean field variables $v_{ij}$, using
the MF equations, the distance measure $d_{ij}^2$ is a fixed quantity.
This corresponds to having the chains at fixed positions. However, we
also want to minimize the distance between the two chains. Based on
the \textit{probabilistic} nature of the mean field variables we
propose to update the chain positions using the (fuzzy) assignment
matrix $\mathcal{V}$, with elements $v_{ij}$. This is done
simultaneously with the updating of $v_{ij}$. Explicitly, one of the chains
will be moved in order to minimize the chain error function $\Echain$
(Eq. (\ref{Echain})). 

The distance measure $d_{ij}^2$ depends on the translation vector
$\mathbf{a}$ and the rotation matrix $\mathcal{R}$, making a total of
six independent variables. Let $\mathbf{x}'_i$ be the coordinates of
the translated and rotated protein, i.e.  $\mathbf{x}'_i = \mathbf{a}
+ \mathcal{R}\mathbf{x}_i$, then
\beq
\label{Echain2}
\Echain = \sum_{i=1}^{N_1}\sum_{j=1}^{N_2} v_{ij} \left(\mathbf{a} +
        \mathcal{R}\mathbf{x}_i - \mathbf{y}_j \right)^2
\eeq
This minimization problem can be solved exactly with 
closed-form expressions for $\mathcal{R}$ and $\mathbf{a}$ that
minimizes $\Echain$ \cite{neumann}. It should be noted that this solution
is rotationally invariant (independent of $\mathcal{R}$) for the special case when the atoms in the two
chains matches each other with the same weight, i.e. when $v_{ij} = 
\mbox{constant}$ for all $i$ and $j$, which is the case for high $T$.

In summary, for a decreasing set of temperatures $T$, one iterates 
until convergence:

\mbox{} 1. The MF equations (Eq. (\ref{MF})).\\
\mbox{} 2. Exact translation and rotation of the chain 
                (Eq. (\ref{Echain2})).

We stress again that step 2 is done with the fuzzy MF assignment 
variables $v_{ij}$ and not with the binary  ones, $s_{ij}$. After 
convergence, $v_{ij}$ are rounded off to $0$ or $1$ and {\it rms} 
(root-mean-square-distance) is computed for the matching pairs. 
Algorithmic details can be found in the next subsection. 

The forces $u_{ij}$ entering Eq. (\ref{MF}) are proportional to
$d_{ij}^2$ (Eqs. (\ref{Echain},\ref{u})). It is the ratio $d_{ij}^2/T$
that counts. Hence, for large temperatures $T$, $v_{ij}$ are fairly
insensitive to $d_{ij}$ and many potential matching pairs ($i,j$)
contribute fairly evenly. As the temperature is decreased, a few pairs
(the ones with small $d_{ij}$) are singled out and finally at the
lowest $T$ only one winner remains. One can view the situation as that
around each atom $i$ one has a Gaussian domain of attraction, which
initially (large $T$) has a large width, but gradually shrinks to a
small finite value.

\begin{figure}[htb]
\begin{center}
{\includegraphics[angle=0,width=14cm]{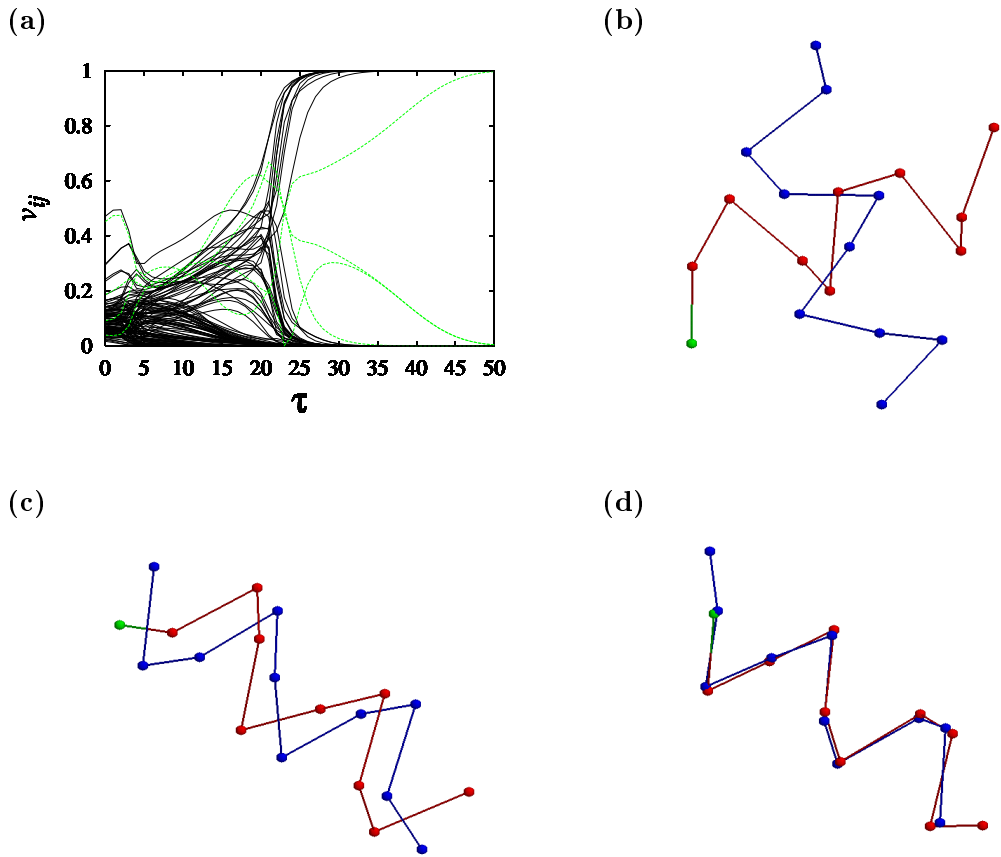}
      }
\end{center}
\caption{ Illustration of the fuzziness of the approach. The alignment shown is 
for 10 atoms in the first helices in the proteins {\sc 1ECD} (blue) and
{\sc 1MBD} (red). {\bf (a)}~Evolution of all the 120 $v_{ij}$ as a
function of iteration time $\tau$ ($T$ is lowered with $\tau$). {\bf
(b)}~Positions of the atoms at $\tau=1$. For high $T$ every atom in a
protein feels all the atoms in the other protein and the problem
is rotationally invariant. {\bf (c)}~$\tau=12$; most of the relevant
matchings are forcing the system to move in the right direction. {\bf
(d)} $\tau=50$; the final assignments are done.  The different
snapshots are presented using different projections. Some $v_{ij}$
approach 0 or 1 rather late and they are coloured green. 
These $v_{ij}$ are related to the atom at the end of the {\sc 1MBD} 
segment, which also is coloured green, and as can be seen in 
{\bf (c)}~ the difficulty is whether to align this atom
to the last or second last atom in the {\sc 1ECD} segment.}
\label{fig_evol}
\end{figure}

The fuzziness of the approach is illustrated in Fig.~\ref{fig_evol},
where the evolution of $v_{ij}$, as $T$ is lowered, is shown for parts
of the first helices of {\sc 1ECD} and {\sc 1MBD} (see next section)
together with snap-shots of the corresponding chain sections. At high
$T$ all $v_{ij}$ are similar; all potential matches have equal
probability. At lower $T$, several $v_{ij}$ have approached $0$ or $1$
and the movable chain is moving in the right direction. At yet lower
$T$, note that a few $v_{ij}$ converge later than the majority. These
are in this example related to the matching of the last atom in one of
the chains. This atom has two potential candidates to match resulting
in a number of $v_{ij}$ that converge last.

\subsection*{Implementation}

Here we give a very condensed, but yet self-contained and detailed
description of the algorithm and the parameters involved, such that 
the results of this paper are reproducible.

{\bf Parameters.} Two kind of parameters are used; the ones related to
encoding of the problem ($\gamma$) and iteration dynamics
($\epsilon$), where $\epsilon$ governs the annealing schedule (see
below), and the ones specifying gap costs ($\lambda$, $\delta$).  The
same set of parameters can be used for most of the pairs (see Table
\ref{param}); the algorithm is remarkably stable.
\begin{table}[ht]
  \begin{center}
    \begin{tabular}{l|cccccc}
      \hline
Protein Family & $\epsilon$ & $\gamma$ &  $\lambda$ & $\lsheet$ & $\lhelix$ & $\delta$  \\
\hline
$\alpha/\beta$, all-$\alpha$ & 0.8 & 0.065 & 0.10 & $1.5\lambda$ & $1.5\lambda$ & $\lambda / 2$ \\    
\hline 
Plastocyanin/azurin & 0.8 & 0.035 & 0.10 & $2.0\lambda$ & $2.0\lambda$ & $\lambda / 5$ \\    
Immunoglobulins     & 0.8 & 0.040 & 0.15 & $2.0\lambda$ & $2.0\lambda$ & $\lambda / 5$ \\    
Winged helix fold   & 0.8 & 0.070 & 0.20 & $2.0\lambda$ & $2.0\lambda$ & $\lambda / 5$ \\    
\hline
    \end{tabular}
  \end{center}
  \caption{Parameters used in the algorithm. The first family involves 27 pairs,
  whereas the others one each.}
  \label{param}
\end{table}

{\bf Initialization.} An initialization of the chains is made
prior to the mean field alignment. First both chains are moved to
their common center of mass. For the \textit{random} initialization, 
this move is then followed by a random rotation of one of
the chains. Most of the times, however, a \textit{sequential} 
initialization is used that consists of minimizing Eq. (\ref{Echain}) using
a band-diagonal assignment matrix $\mathcal{S}$. This corresponds to a
situation where, on the average, atom $i$ in one of the chains is matched
to atom $i$ in the other. If not explicitly mentioned, sequential
initialization is used for all the protein pairs in this paper.

{\bf Iteration Steps.} The shortest chain is always chosen as the
one that is moved ($\mathbf{x}_i$). The mean field variables $v_{ij}$ are
updated according to Eq. (\ref{MF}) where, in order to improve
convergence, the derivatives in Eq. (\ref{u}) are replaced by finite
differences (see e.g. \cite{ohl1}). This update equation accounts for
all mean field variables except for the first row of $\mathcal{V}$,
which is updated according to
\beq
\label{Row0}
v_{0j} = 1 - \sum_{i=1}^{N_1}v_{ij}; \quad j=1,...,N_2
\eeq
The algorithmic steps are shown in Fig.~\ref{MFlg}. After
convergence, no post processing is applied for the results in 
the next section.

\begin{figure}[htb]
  \[\fbox{
    \begin{minipage}[t]{13.0cm}
      \begin{list}{}{\setlength{\itemsep}{2 pt}}
      \item[1.] {Initialization.}
      \item[2.] {Rescale coordinates such that the largest distance
          between atoms within the chains is unity.}
      \item[3.] {Initiate all $v_{ij}$ close to $1/\max(N_1,N_2)$ (randomly).}
      \item[4.] {Initiate the temperature (e.g. $T = 2$).}
      \item[5.] {Randomly (without replacement) select one row, say row $k$.}
      \item[6.] {Update all $v_{kj}, j=0,...,N_2$ according to Eq. (\ref{MF}).}
      \item[7.] {Repeat items $5 - 6$ $N_1$ times (such that all rows have been
          updated once).}
      \item[8.] {Repeat items $5 - 7$ until no changes occur \\
          (defined e.g. by $1/(N_1N_2)\sum_{ij} |v_{ij}-v_{ij}^{\mbox{\scriptsize (old)}}| \leq 0.0001$).}
      \item[9.] {Rotation and translation of the shortest chain using the fuzzy assignment
                 matrix $\mathcal{V}$.}
      \item[10.] {Decrease the temperature, $T \rightarrow \epsilon T$.}
      \item[11.] {Repeat items $5 - 10$ until all $v_{ij}$ are close to 1 or 0 \\
          (defined e.g. by $1/N_1\sum_{ij}v_{ij}^2 \geq 0.99$).}
      \item[12.] {Finally, the mean field solution is given by the 
          integer limit of $v_{ij}$, i.e. \\
          for each row $i$, $i=1,...,N_1$ select the column $j^*$ such that 
          $v_{ij^*}$ is the largest element for this row. Let $s_{ij^*}=1$ and all other
          $s_{ij}=0$ for this row.} 
\end{list}
\end{minipage}
}\]
\caption{Algorithmic steps.}
\label{MFlg}
\end{figure}

%% file: results.tex
To test the quality of our alignment algorithm, we have compared
alignments of protein pairs with results from other automatic
procedures. For most of the tested pairs, each protein belongs to the
same {\sc Scop} superfamily. The goal here is not a full
investigation of all families but rather to explore a limited set with
representative variation.
Pairs were picked from a selection of the families investigated
in~\cite{levitt3}. Our choice of pairs is essentially based on two
criteria. First, the pairs should have diverse structures, and in
particular include all-$\alpha$, all-$\beta$, and $\alpha$/$\beta$
proteins. Second, in~\cite{levitt3} some families are considered to be
{\it very easy}, {\it easy} and {\it difficult} to align, respectively, and we included
pairs from all these categories.
In addition we have tested the algorithm on cases where permutations 
are needed.

Our results are compared with the {\sc Yale Alignment
Server} (http://bioinfo.mbb.yale.\\edu/align/) and {\sc
Dali} (http://www2.ebi.ac.uk/dali/). The {\sc Yale} server
applies post processing to its alignments by removing aligned pairs
with too large root-mean-square-distance ({\it rms}) in an iterative
manner subject to a termination criteria. A similar procedure is of
course possible in our approach, but we have chosen at this stage to
keep the algorithm clean. In the comparisons below we have for the
{\sc Yale} server quoted results both before and after the
post-processing.

\begin{figure}[tb]
  \begin{center}
    \includegraphics[angle=0,width=10cm]{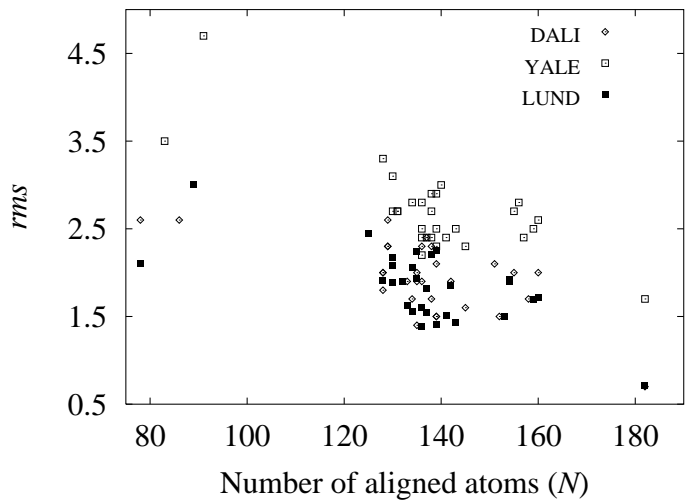}
  \end{center}
  \caption{$rms$ and $N$ corresponding to 
  Table \protect{\ref{t:results}}. The {\sc Yale} data correspond to 
  no post processing (see text).}
  \label{f:N_rms}
\end{figure}

Unless otherwise stated, proteins are in what follows denoted by their
{\sc Pdb}~\cite{pdb} identifier, and in the case of chains or parts of chains with
their {\sc Scop} domain label.
A summary of the results in terms of $rms$ and the number of aligned
atoms~($N$) is shown in Table~\ref{t:results} and in
Fig.~\ref{f:N_rms}.  Detailed comments upon these results and some
additional ones can be found below.

With regard to the general performance one must keep in mind that it
is not straightforward to assess alignment algorithms in terms of e.g.
$rms$ and $N$, since there are no obvious figure-of-merits. It is
interesting to notice though that when inspecting aligned core regions
in detail, we are close to the {\sc Yale} alignments but in general
with a lower $rms$. However, in such comparisons, we differ more from
{\sc Dali}. The {\sc Yale} algorithm has been subject to comparison
with {\sc Scop} classifications using a multiple alignment
procedure~\cite{levitt3}, giving its and our alignments a higher
credibility.

\begin{table}[tb]
\begin{center}
\begin{tabular}{|l|ll@{-\extracolsep{3pt}}l|c|c|c|c|c|c|}
\hline
Protein family & \multicolumn{3}{c}{Protein Pair} \vline & \multicolumn{2}{c}{\sc Yale} \vline & \multicolumn{2}{c}{\sc Dali} \vline & \multicolumn{2}{c}{\sc
Lund} \vline \\ \cline{5-10}
 & \multicolumn{3}{c}{} \vline  & $rms$ & $N$ & $rms$ & $N$ & $rms$ & $N$ \\ \hline 
Dihydrofolate 
& {\sc 1DHFa} & & {\sc 8DFR}  & 1.7 (0.7) & 182 (182) & 0.7 & 182 & 0.7 & 182 \\ 
Reductases 
& {\sc 1DHFa} & & {\sc 4DFRa} & 2.7 (1.2) & 155 (130) & 2.0 & 155 & 1.9 & 154 \\ 
& {\sc 1DHFa} & & {\sc 3DFR}  & 2.5 (1.2) & 159 (143) & 1.7 & 158 & 1.7 & 159 \\ 
& {\sc 8DFR}  & & {\sc 4DFRa} & 2.8 (1.3) & 156 (131) & 2.1 & 151 & 1.9 & 154 \\
& {\sc 8DFR}  & & {\sc 3DFR}  & 2.6 (1.3) & 160 (146) & 2.0 & 160 & 1.7 & 160 \\
& {\sc 4DFRa} & & {\sc 3DFR}  & 2.4 (1.1) & 157 (140) & 1.5 & 152 & 1.5 & 153 \\ \hline
Globins 
& {\sc 2HHBa} & & {\sc 2HHBb} & 2.3 (1.2 )& 139 (129) & 1.5 & 139 & 1.4 & 139 \\  
& {\sc 2HHBa} & & {\sc 2LHB}  & 2.7 (1.6) & 131 (123) & 1.8 & 128 & 1.9 & 130 \\  
& {\sc 2HHBa} & & {\sc 1MBD}  & 2.4 (1.5) & 141 (138) & 1.5 & 139 & 1.5 & 141 \\  
& {\sc 2HHBa} & & {\sc 2HBG}  & 2.4 (0.8) & 138 (105) & 1.7 & 138 & 1.6 & 137 \\  
& {\sc 2HHBa} & & {\sc 1MBA}  & 2.9 (2.2) & 138 (134) & 2.3 & 136 & 2.2 & 138 \\  
& {\sc 2HHBa} & & {\sc 1ECD}  & 3.1 (2.2) & 130 (126) & 2.3 & 129 & 2.2 & 130 \\  
& {\sc 2HHBb} & & {\sc 2LHB}  & 2.5 (1.3) & 136 (126) & 1.7 & 134 & 1.6 & 134 \\  
& {\sc 2HHBb} & & {\sc 1MBD}  & 2.3 (1.4) & 145 (138) & 1.6 & 145 & 1.4 & 143 \\  
& {\sc 2HHBb} & & {\sc 2HBG}  & 2.4 (1.4) & 136 (125) & 2.0 & 135 & 1.6 & 133 \\  
& {\sc 2HHBb} & & {\sc 1MBA}  & 3.0 (2.2) & 140 (137) & 2.3 & 138 & 2.2 & 139 \\  
& {\sc 2HHBb} & & {\sc 1ECD}  & 2.8 (2.2) & 136 (134) & 2.3 & 129 & 2.1 & 134 \\  
& {\sc 2LHB}  & & {\sc 1MBD}  & 2.4 (1.0) & 137 (121) & 1.4 & 135 & 1.4 & 136 \\  
& {\sc 2LHB}  & & {\sc 2HBG}  & 2.7 (1.5) & 131 (119) & 2.0 & 128 & 2.1 & 130 \\  
& {\sc 2LHB}  & & {\sc 1MBA}  & 2.7 (1.8) & 138 (130) & 1.9 & 135 & 1.9 & 132 \\  
& {\sc 2LHB}  & & {\sc 1ECD}  & 2.7 (1.9) & 130 (127) & 2.0 & 128 & 1.9 & 128 \\  
& {\sc 1MBD}  & & {\sc 2HBG}  & 2.5 (1.6) & 139 (130) & 2.1 & 139 & 1.8 & 137 \\  
& {\sc 1MBD}  & & {\sc 1MBA}  & 2.5 (1.7) & 143 (137) & 1.9 & 142 & 1.8 & 142 \\  
& {\sc 1MBD}  & & {\sc 1ECD}  & 2.2 (1.6) & 136 (134) & 1.9 & 136 & 1.6 & 136 \\ 
& {\sc 2HBG}  & & {\sc 1MBA}  & 2.9 (2.2) & 139 (136) & 2.4 & 137 & 2.2 & 135 \\  
& {\sc 2HBG}  & & {\sc 1ECD}  & 3.3 (2.5) & 128 (125) & 2.6 & 129 & 2.4 & 125 \\  
& {\sc 1MBA}  & & {\sc 1ECD}  & 2.8 (1.7) & 134 (125) & 1.9 & 133 & 1.9 & 135 \\ \hline
Plastocyanin/azurin 
& {\sc 1PLC}  & & {\sc 1AZU}  & 4.7 (2.9) & 91  (85)  & 2.6 & 86  & 2.1 & 78  \\ \hline 
Immunoglobulins  
& {\sc 7FABl2}& & {\sc 1REIa} & 3.5 (2.6) & 83  (79)  & 2.6 & 78  & 3.0 & 89  \\ \hline 
\end{tabular}
\caption{ The root-mean-square-distance ({\it rms}) and the number of 
aligned residues ($N$) from the alignment of different  
protein pairs. The results are presented for several automatic 
alignment procedures; {\sc Lund} refers to this work. 
For {\sc Yale} the numbers within parenthesis refer to after post 
processing (see text).}
\label{t:results}
\end{center}
\end{table}
{\bf Dihydrofolate Reductases} ($\alpha/\beta$). These proteins belong
to the {\sc Scop} class $\alpha/\beta$, which contains $\alpha$- and
$\beta$-proteins that have mainly parallel beta sheets.  They are
considered very easy to align~\cite{levitt3}. If we compare
alignments of core structure parts using the three methods we find
that they all essentially agree. However, one notes that the {\sc
Yale} results are very sensitive to the post processing.

{\bf Globins} (all-$\alpha$).  In the all-$\alpha$ {\sc Scop} class we
particularly study a set of globin proteins. In general, we get lower
{\it rms} than the other algorithms for the same number of aligned
residues. When comparing alignments from the three algorithms we find
that an important aspect of our algorithm is manifested -- allowing
for permutation of individual atoms. The reason for this is that to
optimally align secondary structures it is often beneficial to have a
few permuted residues in loops between the secondary structures. If we
again compare the core parts of the alignments from the three
algorithms we find that they agree on a large fraction of the parts.

{\bf Plastocyanin/azurin} (all-$\beta$). All-$\beta$ proteins are
difficult to align if one only takes backbone coordinates ($C_\alpha$
or $C_\beta$) into account, even though using $C_\beta$ instead of
$C_\alpha$ coordinates, in general, improves the results.  As an
initial example of all-$\beta$ proteins we have looked at plastocyanin
versus azurin. Even though this alignment is slightly more difficult
than the previous cases, all three methods give similar $rms$ and $N$
and they all agree on the alignment of a majority of the core parts.
For this example several restarts were performed with random 
initialization.

{\bf Immunoglobulins} (all-$\beta$).  A more difficult example of
all-$\beta$ proteins is immunoglobulins. We align the domain {\sc
  7FABl2} with the chain {\sc 1REIa} and find that we can find
alignments with low $rms$ that look good. However, if we investigate
the alignment in detail we find that atoms in all core regions, except
one, are misaligned. This is also the case in~\cite{levitt3}, where
the same alignment is investigated. To get the core regions correctly
aligned in~\cite{levitt3} they improve their method and take side
chain orientation into account. We expect that this is the case for
our method too. When aligning strands using only $C_\alpha$
coordinates, strands in the two proteins are often matched
satisfyingly to one another while the individual atoms are aligned
such that one strand is translated with respect to the other. It is
therefore obvious that side chain orientation is very important when
aligning strands. For this example several restarts were performed 
with random initialization.

{\bf Permuted proteins -- winged helix fold}.  Finally we look at
permuted versions of similar folds.  We compare two DNA binding
domains related to transcription regulation. The compared domains both
have the winged helix fold but one of them has the secondary
structures in a circularly permuted order. This is a case where
iterative dynamical programming algorithms will fail. We look at {\sc
1LEA} and compare it to the {\sc Entrez-Mmdb}~\cite{entrez_mmdb}
structural domain 4 in chain B of {\sc 1XGN}. This part of {\sc 1XGN}
is classified as a circularly permuted winged helix fold in {\sc
Scop}. In the {\sc Entrez-Mmdb} database, which uses {\sc Vast}
(http://www.ncbi.nlm.nih.gov/Structure/VAST/) for alignments, {\sc
1XGNb4} is listed as a low priority structural neighbour to {\sc
1LEA}, even though {\sc Vast} does not allow for permutations of
secondary structure. If one looks at the actual alignment one finds
that the permuted secondary structures are not aligned. In
Figure~\ref{f:winged_helix_alignment} we compare our alignment with
{\sc Vast}. We show the sequential parts of our alignment and in
particular all parts with secondary structure are shown. {\sc Vast}
aligns only 39 residues in this comparison, while we align 60. We
note that we get all the 39 of the aligned residues of {\sc Vast} but
that we in addition align the sheet at the end of {\sc 1LEA} with the
sheet at the beginning of the domain in {\sc 1XGNb}. This demonstrates
the importance of having a procedure that takes permutations into
account, which our method does.  Otherwise, important similarities
between protein structures will not be found. For this example several
restarts were performed with random initialization.

\begin{figure}[tb]
\begin{center}
\begin{Verbatim}[fontsize=\small,commandchars=\\\{\}]
        
              **************       *******        ******************   
              5           18     26                               59     65    71
              |            |      |                                |      |     |
\textsc{1LEA}          TARQQEVFDLIRDH      PTRAEIAQRLGFRSPNAAEEHLKALARKGVIEIV     -GIRLLQE
\textsc{1XGNb4}        VAQARFLLAKIKRE      FAYRWLQN-D-M-PEGQLKLALKTLEKAGAIYGY     IYMYVRDV
              |            |      |                                |      |     |
            216          229    235                              265    206   212
\end{Verbatim}
\caption{Alignment of {\sc 1LEA} against the {\sc Entrez-Mmdb} domain 
{\sc 1XGNb4}. The '*' denotes atoms also aligned by {\sc Vast}. {\sc
1XGNb4} is a circularly permuted version of {\sc 1LEA} and our method
finds this and aligns the sheet at the end of {\sc 1LEA} with the
sheet at the beginning of the domain in {\sc 1XGNb}.}
\label{f:winged_helix_alignment}
\end{center}
\end{figure}


%% file: discussion.tex
A new approach to structure alignment has been presented and 
explored. It is based upon an error function encoding in terms 
of both binary assignment variables and real-valued atom coordinates. 
The encoding allows for an extensive search through all possible 
alignments, including the ones involving arbitrary permutations.

The error function is efficiently minimized using a mean field 
approximation of the assignment variables and exact
translation/rotation of the atom coordinates. As a by-product 
of this approximation, a probabilistic interpretation of the result 
is available without tedious stochastic simulations. The approach 
is not sensitive to the choice of distance metric, and hence 
to a large extent ignores outliers.

Despite some conceptual similarities with the iterative dynamical 
programming method \cite{levitt}, our approach is probabilistic 
and more general. Also, and maybe more importantly, it is quite 
different since permutations are allowed from the outset. 
For the latter reason, the algorithm in \cite{levitt} cannot 
be derived as a special case in any limit. 

The method is readily extended to handle more detailed chain 
representations (e.g. side-chain orientation) and user-provided constraints 
of almost any kind.

The approach is evaluated using pairs of protein chains  
chosen to represent a wide variety of situations and the  
resulting alignments are successfully compared with other methods 
that are available on WWW-servers. This evaluation is done using 
$C_{\alpha}$-representations of the chains. 

 
Despite being very flexible, generic and covering the entire space of
alignments the method is on the average as fast as \cite{vast},
slightly slower than \cite{levitt} and significantly faster than
\cite{holm}. Also, it is very robust with respect to the algorithmic 
parameters used with a few exceptions. Once side-chains are 
included, the latter will disappear.


\section*{Biological Implications}

There is a strong need for efficient protein structure alignment
algorithms. Aligning proteins forms the basis for studying functional
relationships among proteins and construction of phylogenetic trees.
It is also very important for structure prediction.

\section*{Acknowledgments}
We thank Bo S\"oderberg for valuable suggestions and Guoguang Lu 
for fruitful discussions.
This work was in part supported by the Swedish Natural Science
Research Council and the Swedish Foundation for Strategic Research. 
One of us (CP) thanks the Theory Group at SLAC, where this work 
was initiated, for its hospitality.